# Interface states in $CoFe_2O_4$ spin-filter tunnel junctions

Pavel V. Lukashev,[1†] J. D. Burton,[1] Alexander Smogunov,[2] Julian P. Velev,[3] and Evgeny Y. Tsymbal[1*]

[1]*Department of Physics and Astronomy & Nebraska Center for Materials and Nanoscience, University of Nebraska, Lincoln, Nebraska 68588, USA*

[2] *CEA, Institut Rayonnement Matière de Saclay, SPCSI, F-91191 Gif-sur-Yvette Cedex, France*

[3]*Department of Physics, Institute for Functional Nanomaterials, University of Puerto Rico, San Juan, Puerto Rico 00931, USA*

Spin-filter tunneling is a promising way to generate highly spin-polarized current, a key component for spintronics applications. In this work we explore the tunneling conductance across the spin-filter material $CoFe_2O_4$ interfaced with Au electrodes, a geometry which provides nearly perfect lattice matching at the $CoFe_2O_4$/Au(001) interface. Using density functional theory calculations we demonstrate that interface states play a decisive role in controlling the transport spin polarization in this tunnel junction. For a realistic $CoFe_2O_4$ barrier thickness, we predict a tunneling spin polarization of about -60%. We show that this value is lower than what is expected based solely on considerations of the spin-polarized band structure of $CoFe_2O_4$, and therefore that these interface states can play a detrimental role. We argue this is a rather general feature of ferrimagnetic ferrites and could make an important impact on spin-filter tunneling applications.

In the last few decades spintronics has been one of the most active fields in condensed matter physics, mostly because of its vast potential for device applications.[1] The cornerstone of spintronics is the generation, injection and transport of spin-polarized current (SPC). The conventional approach of manipulating SPC is based on magnetic tunnel junctions (MTJs) in which two ferromagnetic electrodes are separated by a non-magnetic insulating barrier. In MTJs the tunneling current depends on the relative magnetization orientation of the electrodes, effect known as tunneling magnetoresistance (TMR).[2] An alternative approach is to use spin-filter tunneling where a ferro(ferri)magnet is used as a barrier in a tunnel junction with non-magnetic electrodes.[3] Spin-filter tunneling relies on different probabilities for electrons with opposite spin to be transmitted through a spin-dependent energy barrier of the ferro(ferri)magnetic insulator. The spin-dependence of the energy barrier is due to the exchange splitting of the band structure, which leads to the conduction band minimum (CBM) and/or the valence band maximum (VBM) lying at different energies for majority- and minority-spin electrons. The tunneling transmission depends exponentially on the barrier height, therefore tunneling conductance is expected to be spin-dependent.

Despite some promising early experiments on Eu chalcogenides, such as EuS[4] and EuSe[5] and EuO[6], demonstrating the potential of spin-filter tunneling using the Tedrow-Meservey technique[7], practical applications are limited due to their low Curie temperatures. For that reason, the focus recently has shifted to the spinel-based materials, such as $CoFe_2O_4$,[8,9] $NiFe_2O_4$,[10] $NiMn_2O_4$,[11] $BiMnO_3$,[12] $CoCr_2O_4$,[13] and $MnCr_2O_4$,[13] which exhibit much higher Curie temperatures.

The theoretical understanding of the spin-filter tunneling has been largely based on the free-electron model[3,14] and more recently on the analysis of the complex band structure.[15,16] In the former, the spin-filter efficiency is entirely determined by the spin-dependent barrier height in the ferromagnetic insulator. The latter approach takes into account the realistic electronic structure of the bulk material, in particular the orbital character and symmetry of the complex bands. Both approaches work, at best, in the limit of large barrier thickness, thereby neglecting any possible effects of the electrode/barrier interfaces. In particular, the presence of localized interface states are known to play a decisive role in spin-dependent tunneling.[17,18] This question has yet to be addressed for spin-filter systems.

We employ here density functional theory (DFT) calculations to explore spin filtering in a prototype Au/$CoFe_2O_4$/Au (001) tunnel junction. $CoFe_2O_4$ (CFO) has a much narrower minority-spin band gap[9], and hence strong spin filtering with a large negative spin polarization is expected for large thickness of CFO. We demonstrate, however, that majority-spin states present at the $CoFe_2O_4$/Au interface can produce a sizable contribution to the tunneling conductance for reasonable barrier thicknesses (i.e. ~2 nm), thereby reducing the spin polarization anticipated from the complex band structure of bulk CFO alone. We demonstrate that these interface states originate from native surface states of CFO. We argue that such interface states are a rather general feature of ferrimagnetic ferrites and will have an important impact on spin-filter tunneling.

We perform DFT calculations using the Quantum Espresso (QE) package.[19] We use the generalized gradient approximation (GGA) according to the Perdew-Burke-Ernzerhof (PBE) formulation[20] with energy cutoff of 500 eV for the plane-wave expansion and k-point sampling of 6×6×4 (bulk CoFe$_2$O$_4$) and 4×4×1 (heterostructure) for the self-consistent calculations. Tunneling transmission through a CoFe$_2$O$_4$ (CFO) barrier separating two semi-infinite leads of Au is calculated using the wave function-matching formalism implemented for plane waves and pseudopotentials in the QE package.[21,22] All calculations are performed with Hubbard $U$ correction,[23] which is necessary to accurately describe the insulating electronic structure of CFO.[24] We set $U = 3$ eV and $J = 0$ eV for the $d$-orbitals of both Fe and Co, in accordance with a recent theoretical study.[16] Analysis of the complex band structure is achieved by constructing Wannier orbitals from the GGA+$U$ band structure of bulk CFO[25] and using standard tight-binding techniques thereafter.

CFO is a ferrimagnetic insulator with a bulk Curie temperature of 796 K.[9] The oxygen atoms form a face-centered cubic (FCC) sublattice, with cation atoms distributed over tetrahedrally and octahedrally coordinated sites. CFO has an inverse spinel structure with $Fd\bar{3}m$ symmetry with 56 atoms per cell. Fe atoms occupy all of the tetrahedral whereas the octahedral sites are randomly occupied by Co and Fe.[24] For a manageable computational cell we arrange the Co and Fe atoms on the octahedral sites in order to increase the symmetry of the cell. This allows a reduction in the size of the unit cell to a tetragonal cell of 28 atoms with space group $Imma$. In this geometry the calculated lattice parameters for the CFO are $a = 5.91$ Å and $c/a = 1.41$.

The ground state of CFO is ferrimagnetic, where magnetic moments on octahedral sites are aligned parallel to one another, but antiparallel to the magnetic moments of Fe atoms at tetrahedral sites. The magnetic moments projected on individual atomic sites are 2.5 $\mu_B$ for Co, 4.0 $\mu_B$ for Fe at octahedral sites, and $-3.9$ $\mu_B$ for Fe at tetrahedral sites. There are also induced magnetic moments on O atoms: 0.05 $\mu_B$ per O in the CoO$_2$ planes and 0.15 $\mu_B$ per O in the FeO$_2$ planes. The total magnetic moment of CFO is 3.0 $\mu_B$ per formula unit, consistent with the expected formal electronic configurations of the transition-metal cations (Co$^{2+}$ and Fe$^{3+}$ both in their high-spin configurations) and the ferrimagnetic alignment.

Fig. 1 (b) shows the calculated local densities of states (LDOS) for the bulk CFO. We find that a band gap is about 0.8 eV, determined by minority-spin states, consistent with previous DFT+$U$ calculations of CFO which report band gaps in the range of 0.5 to 1 eV.[24,26,16] The exchange splitting of the CBM is $\Delta_{ex} = 0.9$ eV, consistent with previously reported values in the range of 0.5 to 1.2 eV.[24,26,16] The VBM is predominantly composed of Co (hybridized with O) states, while the CBM in both spin channels are composed of Fe states. Thus, $\Delta_{ex}$ is almost entirely due to the splitting between the Fe states on the octahedral and tetrahedral sites, in agreement with recently published data.[16]

Fig. 1 (a) shows an Au/CFO/Au supercell used in our calculations. We construct the supercell by lattice matching (001) oriented $fcc$ Au with bulk CFO, leading to a tensile strain on the Au of less than 1%. We assume a CoO$_2$ termination of the CFO (001) layer and place interfacial Au atop O atoms. The supercell contains 8 formula units of CFO plus an additional monolayer (ML) of Co$_2$O$_4$ to ensure symmetric interface termination, resulting in non-stoichiometry of the CFO barrier. The structure is then fully optimized with constrained in-plane lattice parameter of bulk CFO, $a = 5.91$ Å.

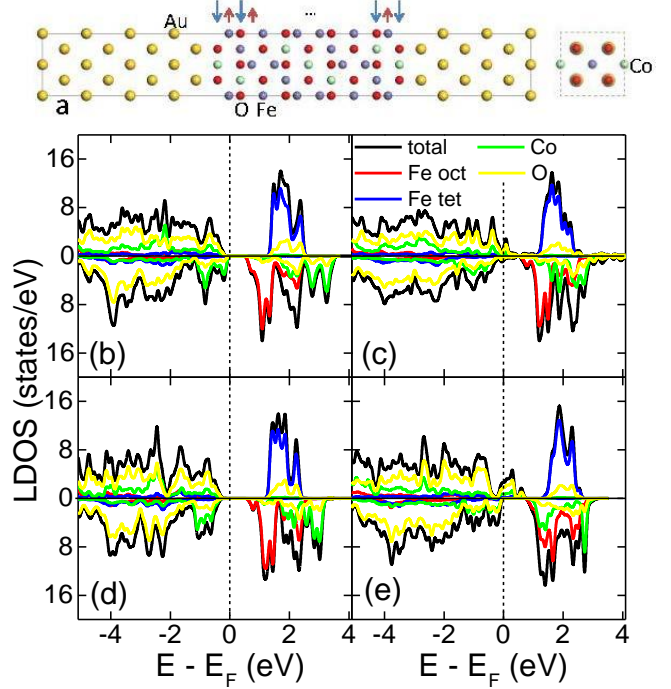

Fig. 1: (color online) (a) Structural model for the Au/CFO/Au tunnel junction. The magnetic moment direction in each layer is indicated by the arrows. LDOS of (b) bulk CFO, (c) interfacial and (d) middle CFO layers of the Au/CFO/Au tunnel junction, and (e) the surface layer of stand-alone (001) CFO slab. Green line – octahedral Co, red line – octahedral Fe, blue line – tetrahedral Fe, yellow line – O, black line – total LDOS. The majority- and minority-spin LDOS are displayed in the upper and lower panels, respectively. The vertical dashed lines denote the Fermi energy.

The calculated LDOS for the Au/CFO/Au tunnel junction is shown in Fig.1 for interfacial (Fig. 1 (c)) and middle (Fig. 1 (d)) CFO layers. While the LDOS for the middle CFO layer closely resembles that of bulk (compare Fig. 1 (b) and (d)), the interface LDOS exhibits different behavior.[27] As is evident from Fig. 1(c), interface states appear within the band gap of CFO for the majority-spin electrons, with a peak near the Fermi energy ($E_F$).

The CFO/Au interface states originate from native CFO (001) surface states, as confirmed from a separate calculation of a stand-alone CFO (001) slab with the same structure as in the supercell. The surface LDOS of this slab (Fig. 1(e)) displays surface states in the bulk gap for majority- but not for minority-spins. Details of the surface states are shown in Fig. 2. In Fig. 2(a-c) the $\mathbf{k}_\parallel$-resolved LDOS is plotted in the two-dimensional Brillouin zone (2D BZ) for the surface atomic layer in the CFO (001) slab, calculated for the majority-spin at different energies (see the figure caption). Fig. 2(d) shows the majority-spin density for the (001) CFO slab calculated by integrating the surface layer LDOS from $E_F$ to $E_F + 0.4$ eV. The majority-spin surface states mostly consist of O-$p_x$, O-$p_y$, and Co-$d_{xy}$ orbitals, as shown on Fig. 2(d) and confirmed by additional calculations of the orbitally-resolved $\mathbf{k}_\parallel$-distribution (see Supplementary Materials). These states originate from the fact that, at the surface, the Co atoms lose their octahedral coordination due to one "missing" O atom at the apex. The octahedral crystal field in the bulk splits the Co $d$-states into a low energy $t_{2g}$ and higher energy $e_g$ manifold. Absence of the apex O atom at the surface further splits the $e_g$ states which make up the majority-spin VBM, lowering the $d_{z^2}$ states and raising the $d_{xy}$ states.[28] The higher crystal field of the $d_{xy}$ orbitals leads to the formation of the surface states. As seen in Fig. 2(d), the structure consists of relatively well-separated parallel chains of $CoO_2$ oriented along the $y$-direction, leading to larger dispersion along $y$ than $x$ and therefore giving rise to the two-fold rotational symmetry seen in Fig. 2(a-c).

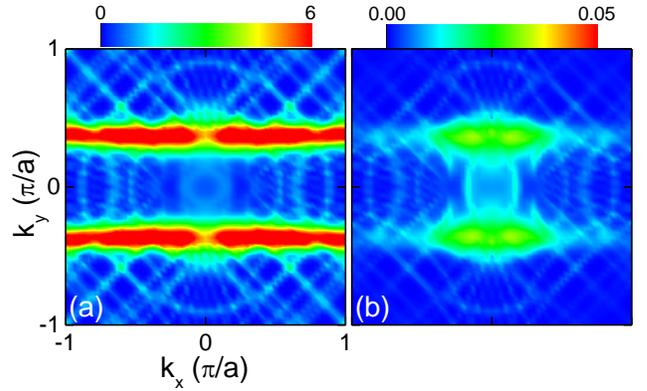

**Fig. 3:** (color online) $\mathbf{k}_\parallel$-resolved majority-spin LDOS (arbitrary units) at the Fermi energy for (a) interfacial and (b) middle CFO layers in Au/CFO/Au tunnel junction.

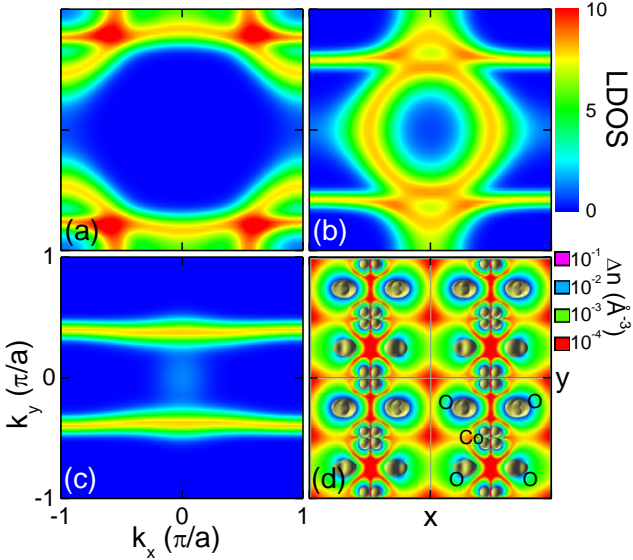

**Fig. 2:** (color online) (a-c) $\mathbf{k}_\parallel$-resolved majority-spin LDOS (arbitrary units) at (a) $E_F$, (b) $E_F + 0.2$ eV and (c) $E_F + 0.4$ eV for the surface atomic layer in the stand-alone (001) CFO slab. (d) Integrated LDOS, $\Delta n$, in real space for the surface layer of the (001) CFO slab from $E_F$ to $E_F + 0.4$ eV. Color indicates the density on a plane cutting through the surface Co atoms and the shaded surfaces correspond to a constant-density $\Delta n = 0.05$ Å$^{-3}$.

These majority-spin surface states survive at the Au/CFO interface, as can be seen in the $\mathbf{k}_\parallel$-resolved LDOS for the interfacial CFO layer, plotted in Fig. 3. As seen from Fig. 3 (a), these states exhibit the same distinct stripe-like features originating from the CFO surface states (compare Fig. 2(c) and Fig. 3(a)). We note that the interface states shown of Fig. 3(a) are calculated at $E_F$ of the Au/CFO/Au tunnel junction, which is shifted by about 0.25 eV away from the VBM in the (001) CFO slab. The surface states shown on the Fig. 2(c) are plotted at $E_F + 0.4$ eV. The small energy difference is due to the slightly different nature of the LDOS at $E_F$ for the CFO surface and interface.[29] These majority-spin interface states of CFO in Au/CFO/Au tunnel junction have a significant effect on the tunneling conductance with CFO, as confirmed below.

We calculate tunneling conductance by taking the Au/CFO/Au supercell as a scattering region and attaching it on both sides to semi-infinite FCC Au leads. The calculations are performed at zero bias using a uniform 60 × 60 k-point mesh in the 2D BZ. The calculated conductance per unit cell area is $G_\uparrow = 0.11 \times 10^{-4}$ $e^2/h$ for majority- and $G_\downarrow = 0.40 \times 10^{-4}$ $e^2/h$ for minority-spin channels, respectively. The spin polarization of the tunneling current is $P = (G_\uparrow - G_\downarrow)/(G_\uparrow + G_\downarrow) = -57\%$. The negative sign of $P$ is consistent with the expectation following from the lower minority-spin band gap compared to the majority-spin band gap and is in agreement with experimental results for fully epitaxial junctions with CFO as a tunneling barrier.[9]

Figs. 4(a-b) show the $\mathbf{k}_\parallel$- and spin-resolved conductance of the Au/CFO/Au tunnel junction. The majority-spin conductance, Fig. 4(a), can be explained by correlating it with the $\mathbf{k}_\parallel$-resolved LDOS shown in Fig. 3 (b). The interface states seen in Fig. 3(a) as stripes for the interfacial CFO layer strongly decay away from the interface, however, even in the middle of the CFO barrier layer they do not completely vanish (Fig. 3 (b)). Moreover, comparison of Fig. 4 (a) with Fig. 3 (b) indicates a clear correlation between the $\mathbf{k}_\parallel$- resolved conductance and LDOS profiles, both exhibiting maxima in the same area of the 2D BZ. The transmission distribution bears little resemblance, however, to the distribution of lowest decay rates for majority spins, Fig. 4(c), as determined by the complex band structure calculations. We conclude, therefore, that the tunneling conductance of majority-spin electrons is, in fact, dominated by the interface states and therefore cannot be deduced by consideration of the complex band structure of CFO alone.

The conductance profile for the minority-spins, on the other hand, is very reminiscent of the distribution of evanescent states in the band gap of CFO. Figure 4(d) shows the lowest decay rates of the minority-spin evanescent states in the band gap of CFO, where we see a close resemblance between the conductance (Fig. 4(b)) and the decay rate distribution (Fig. 4(d)) for the minority-spin. Finally, we notice that both majority- and minority-spin channels demonstrate minimal conductance at the Γ point (Fig. 4 (a-b)), somewhat inconsistent with the distribution of the decay rates (Fig. 4 (c-d)) and with recently published results.[16] This is due to the mismatch of the band symmetries for both majority and minority spin channels of Au and CFO, calculated for $k_x = k_y = 0$, along the [001] direction. In particular, for Au along this direction there is only one band crossing Fermi level having $\Delta_1$ symmetry, i.e. with orbital contributions $s$, $p_z$, and $d_{z^2}$. None of these orbital characters belong to the slowest decaying bands near the VBM of CFO, being primarily of $p_y$ and $d_{xy}$ orbital character for majority-spin and $d_{x^2-y^2}$ and $p_y$ for minority-spin.

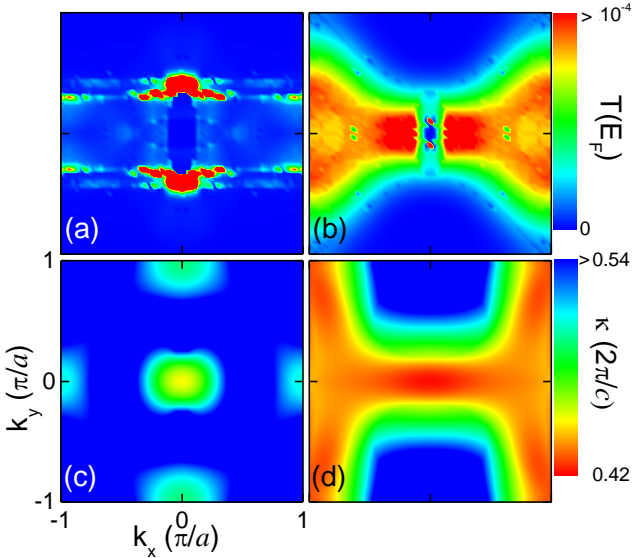

**Fig. 4:** (color online). $\mathbf{k}_\parallel$-resolved transmission at $E_F$ for (a) majority- and (b) minority-spin channels of the Au/CFO/Au tunnel junction. Lowest decay rate, κ, of the (c) majority- and (d) minority-spin evanescent states of bulk CFO as a function of $\mathbf{k}_\parallel$ in the 2D BZ at VBM + 0.4 eV.

The contribution of the majority-spin interface states is detrimental to the net spin-polarization of the tunneling conductance. To see this, we return to the simpler description of the spin-filter effect based solely on the complex band structure where we assume featureless electrodes and perfect interface transmission functions. In this case the conductance for each spin-channel is determined by $G \propto \int e^{-2\kappa(\mathbf{k}_\parallel)t} d^2\mathbf{k}_\parallel$ where $t$ is the thickness of the barrier, $\kappa(\mathbf{k}_\parallel)$ is the calculated lowest decay rate at $E_F$ and $\mathbf{k}_\parallel$ and the integral is over the entire 2D BZ. Using $t = 1.9$ nm and $E_F$ = VBM + 0.4 eV, we find a spin-polarization of $P = -80\%$. This is significantly larger than what is found from our full transport calculations, where interface states dominate the majority spin channel.

The predicted effect of interface states on spin-polarized tunneling is not limited to the particular geometry of the tunnel junction considered above. We find that a terminating layer of the CFO (001) with a mixture of Fe and Co, as well as a purely $FeO_2$ terminating layer, both also lead to majority-spin interface states which produce similar detrimental effects on spin-polarized tunneling. One could expect a different behavior for Fe at tetrahedral sites comprising the interface; we find, however, that this termination is unstable.

In summary, we have shown that the spin polarization of the tunneling conductance in Au/$CoFe_2O_4$/Au (001) tunnel junction is strongly affected by majority-spin interface states, leading to a reduction in spin-polarization as compared to expectations based on the spin-polarized band-gap alone. Interface states are a general feature of the ferrimagnetic ferrites that are used as spin-filter barriers. Thus, the predicted effect has important implications for the design of spin-filter tunnel junctions, where the interface states need to be avoided to exploit the unspoiled spin filtering anticipated from the band structure of the bulk material.

This research was supported by the NSF (Grant No. EPS-1010674) and the Nebraska Research Initiative. The work at the University of Puerto Rico was supported by NSF Grants DMR-1105474 and EPS-1010094. Computations were performed at the University of Nebraska, Holland Computing Center.

† pavel.lukashev@unl.edu
* tsymbal@unl.edu

1. *Handbook of Spin Transport and Magnetism*, eds. E. Y. Tsymbal and I. Žutić (CRC press, Boca Raton, FL, 2011), 808 pp.
2. E. Y. Tsymbal, O. N. Mryasov, and P. R. LeClair, *J. Phys.: Cond. Matt.* **15**, R109 (2003).
3. T. S. Santos and J. S. Moodera, in *Handbook of Spin Transport and Magnetism in Electronic Systems*, eds. E. Y. Tsymbal and I. Žutić, Ch. 13 (CRC Press, Boca Raton, FL, 2011), p. 251.
4. J. S. Moodera, X. Hao, G. A. Gibson, and R. Meservey, *Phys. Rev. Lett.* **61**, 637 (1988).
5. J. S. Moodera, R. Meservey, and X. Hao, *Phys. Rev. Lett.* **70**, 853 (1993).
6. T. Santos and J. Moodera, *Phys. Rev. B* **69**, 241203(R) (2004).
7. R. Meservey and P. M. Tedrow, *Phys. Reports* **238**, 173 (1994).
8. M. G. Chapline and S. X. Wang, *Phys. Rev. B* **74**, 014418 (2006).
9. A. V. Ramos, M. J. Guittet, J. B. Moussy, R. Mattana, C. Deranlot, F. Petroff, and C. Gatel, *Appl. Phys. Lett.* **91**, 122107 (2007).
10. U. Lüders, M. Bibes, K. Bouzehouane, E. Jacquet, J.-P. Contour, S. Fusil, J.-F. Bobo, J. Fontcuberta, A. Barthélémy, and A. Fert, *Appl. Phys. Lett.* **88**, 082505 (2006).
11. B. B. Nelson-Cheeseman, R. V. Chopdekar, L. M. B. Alldredge, J. S. Bettinger, E. Arenholz, and Y. Suzuki, *Phys. Rev. B* **76**, 220410(R) (2007).
12. M. Gajek, M. Bibes, A. Barthélémy, K. Bouzehouane, S. Fusil, M. Varela, J. Fontcuberta, and A. Fert, *Phys. Rev. B* **72**, 020406(R) (2005).

13. R. V. Chopdekar, B. B. Nelson-Cheeseman, M. Liberati, E. Arenholz, and Y. Suzuki, *Phys. Rev. B* **83**, 224426 (2011).
14. M. Müller, G.-X. Miao, and J. S. Moodera, Europhys. Lett. **88**, 47006 (2009); Phys. Rev. Lett. **102**, 076601 (2009).
15. P. V. Lukashev, A. L. Wysocki, J. P. Velev, M. van Schilfgaarde, S. S. Jaswal, K. D. Belashchenko, and E. Y. Tsymbal, Phys. Rev. B **85**, 224414 (2012).
16. N. M. Caffrey, D. Fritsch, T. Archer, S. Sanvito, and C. Ederer, Phys. Rev B **87**, 024419 (2013).
17. E. Y. Tsymbal, K. D. Belashchenko, J. Velev, S. S. Jaswal, M. van Schilfgaarde, I. I. Oleynik, and D. A. Stewart, Prog. Mater. Sci. **52**, 401 (2007).
18. J. P. Velev, P. A. Dowben, E. Y. Tsymbal, S. J. Jenkins, and A. N. Caruso, Surf. Sci. Rep. **63**, 400 (2008).
19. P. Giannozzi *et al.*, J. Phys.: Condens. Matter **21**, 395502 (2009).
20. J. P. Perdew, K. Burke, and M. Ernzerhof, Phys. Rev. Lett. **77**, 3865 (1996).
21. H. J. Choi and J. Ihm, Phys. Rev. B **59**, 2267 (1999).
22. A. Smogunov, A. Dal Corso, and E. Tosatti, Phys. Rev. B **70**, 045417 (2004).
23. A. I. Liechtenstein, V. I. Anisimov, and J. Zaanen, Phys. Rev. B **52**, R5467 (1995).
24. D. Fritsch and C. Ederer, Phys. Rev. B **82**, 104117 (2010).
25. D. Korotin, A. V. Kozhevnikov, S. L. Skornyakov, I. Leonov, N. Binggeli, V. I. Anisimov, and G. Trimarchi, The European Physical Journal B **65**, 91 (2008).
26. Z. Szotek, W. M. Temmerman, D. Ködderitzsch, A. Svane, L. Petit, and H. Winter, Phys. Rev. B **74**, 174431 (2006).
27. By symmetry DOS are slightly different at right and left CFO interfaces, but the difference is small and ignored throughout the text.
28. Here $x$ and $y$ to refer to the in plane lattice vectors. These are rotated by 45° with respect to the usual coordinate system used to describe octahedrally coordinated transition metals. Therefore, in our notation, $d_{xy}$ belongs to the $e_g$ manifold, with lobes along the directions of O nearest neighbors as seen in Fig. 2(d).
29. We confirmed the presence of the CFO surface states by an additional calculation for a Au/CFO/Au tunnel junction with different interface termination. In particular, similar stripe-like features are present for $\mathbf{k}_\parallel$- resolved DOS of the interfacial CFO layer if the interface layer of CFO consists of both Co and Fe.

# Interface states in $CoFe_2O_4$ spin-filter tunnel junctions: supplementary materials

Pavel V. Lukashev,[1][†] J. D. Burton,[1] Alexander Smogunov,[2] Julian P. Velev,[3] and Evgeny Y. Tsymbal[1][*]

[1]*Department of Physics and Astronomy & Nebraska Center for Materials and Nanoscience, University of Nebraska, Lincoln, Nebraska 68588, USA*

[2]*CEA, Institut Rayonnement Matière de Saclay, SPCSI, F-91191 Gif-sur-Yvette Cedex, France*

[3]*Department of Physics, Institute for Functional Nanomaterials, University of Puerto Rico, San Juan, Puerto Rico 00931, USA*

## Orbital resolved surface states in $CoFe_2O_4$

Fig. S1(a-c) shows the orbital-resolved contributions to the majority-spin surface states calculated at $E_F + 0.3$ eV for the (001) $CoFe_2O_4$ (CFO) slab ($E_F$ – Fermi energy). We see that the majority-spin surface states mostly consist of O-$p_x$, O-$p_y$, and Co-$d_{xy}$ orbitals. All other contributions are negligibly small and are not shown here.

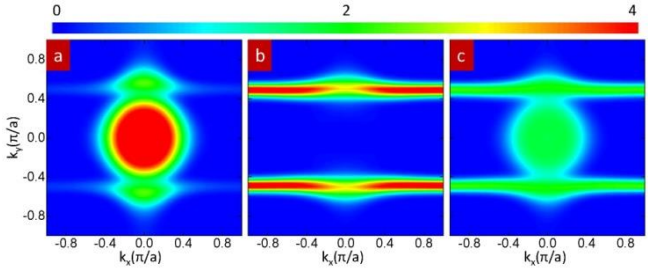

**Fig. S1:** (color online) $k_∥$-resolved majority-spin DOS (arbitrary units) of (001) CFO slab calculated at $E_F + 0.3$ eV for O-$p_x$ (a), O-$p_y$ (b), and Co-$d_{xy}$ (c) orbitals.

## Complex band structure of CFO

Fig. S2 shows the calculated spin-dependent complex band structure of CFO for $k_∥ = 0$ in the $\Gamma \to Z$ direction. The complex bands (left and right panels) are connected to the real bands (middle panel) and inherit their symmetry properties. The curvature for complex and real bands is the same at the connecting points due to the analytic properties of the energy dispersion function, $E(k_z)$. For detailed discussion of the complex band structure's significance for the spin-filter materials, see Ref. [1].

1. P. V. Lukashev, A. L. Wysocki, J. P. Velev, M. van Schilfgaarde, S. S. Jaswal, K. D. Belashchenko, and E. Y. Tsymbal, Phys. Rev. B **85**, 224414 (2012).

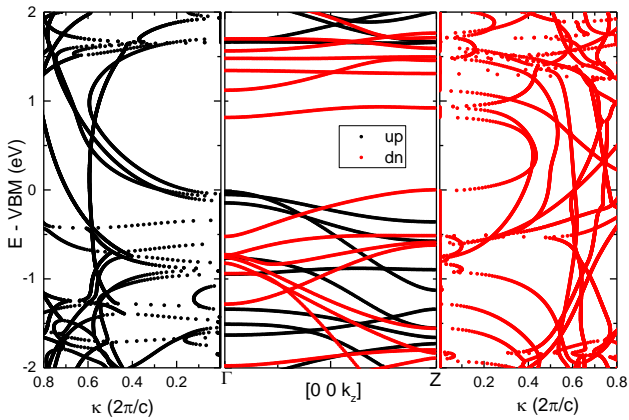

**Fig. S2:** (color online) Complex band structure of CFO in the $\Gamma \to Z$ direction for majority (left panel) and minority (right panel) spin. The middle panel shows real bands for the same direction.

1